\documentclass[a4paper,12pt]{article}
\usepackage{graphicx}
\usepackage{amsmath}
\usepackage[cp1251]{inputenc}
\usepackage[russian,english]{babel}

\begin{document}
\selectlanguage{russian}
\begin{otherlanguage}{english}
\begin{center}
{\bf Ideal synchronizer for marked pairs in fork-join network}
\\
\bigskip

S.~V.~Vyshenski\dag\footnote[1]{svysh@pn.sinp.msu.ru},
P.V.~Grigoriev\ddag, and
Yu.Yu.~Dubenskaya\S
\\
\bigskip

\dag~Institute of Nuclear Physics, Moscow State University, Moscow 119899,
Russia \\
\ddag~General Physics Institute, Russian Academy of Sciences, Vavilov str., 38,
Moscow 119991, Russia\\
\S~Institute of Precise Mechanics and Computer Engineering, Russian Academy of Sciences,
Leninskiy av., 51, Moscow 119991, Russia
\end{center}

\bigskip

\begin{abstract}
We introduce a new functional element (synchronizer for marked pairs)
meant to join results of parallel processing in
two-branch fork-join queueing network. Approximations for distribution of
sojourn time at the synchronizer are derived along with
a validity domain. Calculations are performed assuming that:
arrivals to the network form a Poisson process,
each branch operates like an M/M/N queueing system.
It is shown that mean sojourn time at a real synchronizer node
is bounded below by the value, defined by
parameters of the network (which contains the synchronizer)
and does not depend upon performance and particular properties
of the synchronizer.

\end{abstract}

\end{otherlanguage}

\bigskip
\pagebreak

\begin{center}
{\bf Идеальный синхронизатор маркированных пар
в сети разветвление-объединение
}
\bigskip

С.В.~Вышенский\dag$^1$,
П.В.~Григорьев\ddag,
Ю.Ю.~Дубенская\S
\\
\bigskip

\dag~НИИ ядерной физики МГУ, Москва 119899\\
\ddag~Институт общей физики РАН, Вавилова 38, Москва 119991\\
\S~Институт точной механики и вычислительной техники РАН, Ленинский просп. 51, Москва 119991

10 января 2008 г.

Статья публикуется в журнале\\
\emph{Обозрение прикладной и промышленной математики, 2008}

\end{center}

\begin{abstract}
Предложен функциональный элемент (синхронизатор маркированных пар)
для объединения результатов параллельной обработки двух потоков в сетях
массового обслуживания типа разветвление-объединение (fork-join).
Получены приближения для распределения времени пребывания заявки
в синхронизаторе. Найдена область применимости приближений.
Расчеты проведены для стационарного режима в следующих предположениях:
на вход сети поступает поток заявок пуассоновского типа,
системы в обеих ветвях сети относятся к типу $M/M/N$.
Показано, что среднее время пребывания заявки
в реальном синхронизаторе ограничено снизу значением,
которое определяется параметрами сети, содержащей синхронизатор,
и не зависит от производительности и особенностей синхронизатора.

\end{abstract}

\tableofcontents
\pagebreak

\section{Введение}

\begin{figure}[htb]
\begin{center}
\includegraphics[width=10cm]{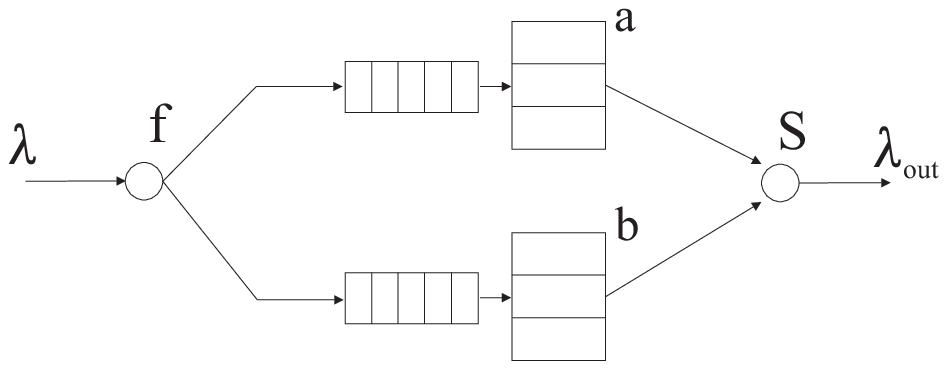}
\end{center}
\caption{
Сеть разветвление-объединение с двумя ветвями.}
\label{PictForkJoinDouble}
\end{figure}

Рассмотрим сеть
массового обслуживания типа разветвление-объединение (РО) или fork-join,
показанную на рисунке \ref{PictForkJoinDouble}.
Предположим, что на вход сети поступает пуассоновский поток промаркированных
(например, пронумерованных)
заявок с интенсивностью $\lambda > 0$.
В точке разветвления $f$ каждая заявка разделяется на две заявки с одинаковыми номерами,
совпадающими с номером исходной заявки.
Эти две заявки одновременно поступают на вход ветвей $a$ и $b$,
которые представляют собой системы массового обслуживания
$M/M/N_a$ и $M/M/N_b$, где $N_a, \,N_b \geq 1$ задают количества
параллельных каналов обслуживания в ветвях $a$ и $b$.
Очереди в ветвях $a$ и $b$ подчиняются дисциплине FIFO.
Сеть функционирует в стационарном режиме.
Для объединения результатов параллельной обработки двух потоков в
ветвях $a$ и $b$ сети служит узел $S$, названный нами
\emph{синхронизатором маркированных пар}.

Сеть массового обслуживания РО представляет собой удобную модель для расчета
нагрузочных характеристик фрагментов различных информационных,
коммуникационных и производственных систем.
В частности, такой фрагмент характерен для систем, функциональность которых
принято задавать на языке потоков работ (workflow).

\begin{figure}[htb]
\begin{center}
\includegraphics[width=7cm]{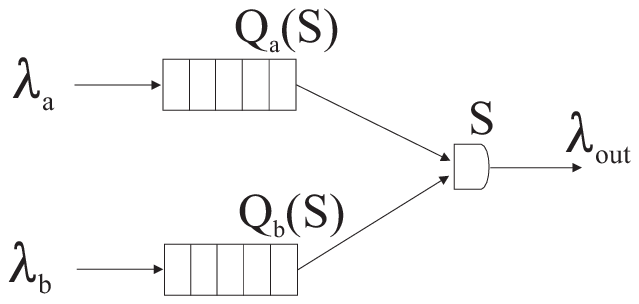}
\end{center}
\caption{
Синхронизатор немаркированных пар в сети типа сборки.
}
\label{PictAssemblyLikeDouble}
\end{figure}

Первоначально проблема описания процесса
объединения двух потоков заявок была изучена на примере
другой сети массового обслуживания - сети типа сборки (assembly-like),
показанной на рисунке \ref{PictAssemblyLikeDouble}.
В такой сети объединяются пары из произвольной заявки типа $a$ и произвольной
заявки типа $b$. Будем называть такие пары \emph{немаркированными}.
В работах \cite{Harrison1973,Latouche1981} показано,
что сеть типа сборки неустойчива, если
объединяемые потоки полностью независимы друг от друга.
В работе \cite{Bonomi1987} (по-видимому, впервые) функциональный элемент $S$,
реализующий объединение потоков заявок в сети типа сборки, назван синхронизатором.
Будем называть такой элемент \emph{синхронизатором немаркированных пар},
чтобы подчеркнуть отличие от синхронизатора маркированных пар,
рассмотренного в настоящей работе.

Сеть типа РО является
развитием сети типа сборки.
От сети типа сборки ее отличает наличие статистической зависимости потоков заявок
в параллельных ветвях (то есть потоков, порожденных в прошлом общим событием - разветвлением,
и подлежащих объединению в будущем), а также то обстоятельство,
что здесь рассматривается объединение маркированных заявок.

В большинстве работ о сетях РО решается задача вычисления
времени отклика сети, то есть промежутка времени, разделяющего моменты разветвления
и объединения потоков заявок в ветвях сети.
В одной из первых работ, Флатто и Хана \cite{Flatto1984}, найдена производящая функция
для стационарного распределения в сети с двумя ветвями $M/M/1$,
а также приведены оценки производительности сети.

В работах Бачелли \cite{Baccelli1985,Baccelli1989},
а также Рагавана и Висванадама \cite{Raghavan2001} найдены ограничения для
времени отклика для различных сетей РО с произвольным
временем обслуживания.
Нельсон и Тантави \cite{Nelson1988} предложили метод аппроксимации времени отклика сети с
ветвями $M/M/1$ и одинаковой интенсивностью обслуживания в ветвях. Они также
использовали производящие функции для получения точного значения
времени отклика в сети с двумя ветвями $M/M/1$.
Варна и Маковски \cite{Varma1994}
предложили ряд эвристических приближений для расчета времени отклика в сетях с произвольным
распределением входного потока и произвольным временем обслуживания.

Для сетей РО с произвольным временем обслуживания Айан и Сео \cite{Ayhan2001}
нашли преобразование Лапласа для времени отклика, а также привели аппроксимацию времени отклика
в случае невысокой входной нагрузки.
В работах Кнессла \cite{Knessl1991}, Кушнера \cite{Kushner2001}, Варна и Маковски \cite{Varma1994}
и Нгуена \cite{Nguyen1993} для оценки параметров сети РО в случае
высокой входной нагрузки применяется диффузионное приближение.

В работе Ко и Серфозо  \cite{Serfozo2004} предложено приближенное выражение для функции распределения
времени отклика в сети РО с $m$ ветвями $M/M/N$.
Там же найдено приближение для функции распределения для общего количества заявок в сети с $m$ ветвями $M/M/1$.

В процитированных выше работах не конкретизировался механизм объединения потоков заявок и
не исследовались статистические свойства этого процесса.

В настоящей работе предлагается явно выделить и исследовать синхронизатор маркированных пар
как отдельный функциональный элемент сети РО,
а также получить
выражения для функции распределения времени пребывания
заявок в идеальном (бесконечно быстром) синхронизаторе.
Такой функциональный элемент по отношению
к сети РО будет играть роль аналогичную той роли, которую играл синхронизатор
немаркированных пар по отношению к сети типа сборки.

Дальнейшее изложение построено следующим образом.
В разделе \ref{SectModels} приведен общий алгоритм работы идеального синхронизатора.
В разделе \ref{SectTime} для сети РО при $N_a = N_b = 1$
путем численного решения уравнений
Колмогорова-Чепмена исследована корреляция между временами
пребывания заявок в ветвях $a$ и $b$.
Для сети РО при $N_a = N_b = \infty$ аналитически получено точное
выражение для распределения времени пребывания
заявки в синхронизаторе.
При конечных значениях $N_a, \,N_b \geq 1$ получены приближенные
выражения для распределения времени пребывания
заявки в синхронизаторе.
Качество предложенных приближений исследовано с помощью
статистического моделирования.
Показано, что даже в идеальном синхронизаторе
среднее время пребывания заявок в синхронизаторе ограничено снизу значением,
которое определяется лишь свойствами сети РО, содержащей синхронизатор,
и не зависит от свойств самого синхронизатора.
В разделе \ref{SectDiscus} полученные в настоящей статье результаты
сравниваются с известными из литературы, а также обсуждаются возможные
применения.
В Приложении \ref{SectForm} найдены приближенные выражения для функции распределения
времени пребывания в синхронизаторе при конкретных значениях параметров $N_a$ и $N_b$.
В Приложении \ref{SectKolm} изложен использованный метод численного решения уравнений
Колмогорова-Чепмена в стационарном случае для сети РО при $N_a = N_b = 1$.
В Приложении \ref{SectStat} изложены использованные методы численного моделирования
для проверки статистических гипотез.

\section{Идеальный синхронизатор маркированных пар} \label{SectModels}

Процесс объединения двух потоков маркированных заявок в разных сетях может происходить
по-разному. В настоящей работе мы предлагаем приближенное описание этого процесса
с помощью нового функционального элемента -- синхронизатора
$S$ маркированных пар (рис. \ref{PictForkJoinDouble}).
Будем считать, что этот синхронизатор $S$ состоит из
памяти синхронизатора и монитора,
который отслеживает и изменяет состояние памяти синхронизатора.
Полагаем, что монитор срабатывает мгновенно,
а размер памяти неограничен.
Именно поэтому мы называем синхронизатор \emph{идеальным}.
Идеальный синхронизатор работает следующим образом.

Заявки из ветвей $a$ и $b$ поступают на вход синхронизатора $S$ маркированных пар
и сохраняются в его памяти.
Пару заявок различных типов ($a$ и $b$) с одинаковыми номерами будем
называть \emph{партнерами}.
Партнера, достигшего синхронизатора первым из пары партнеров, будем называть \emph{первым партнером}.
Партнера, достигшего синхронизатора вторым из пары партнеров, будем называть \emph{вторым партнером}.

После сохранения вновь поступившей заявки в памяти, монитор производит
следующие действия:

\begin{itemize}
\item определяется номер и тип ($a$ или $b$) новой заявки,
\item если в памяти синхронизатора уже имеется (первый) партнер для новой заявки,
то оба партнера найденной пары удаляются из памяти
синхронизатора и передаются на выход синхронизатора.
\item если партнер для новой заявки в памяти отсутствует, то новая заявка
(то есть первый партнер из пары)
остается в памяти синхронизатора ждать своего второго партнера.
\end{itemize}

Исходя из приведенного алгоритма работы синхронизатора, определим
время $t$ пребывания пары в синхронизаторе как разницу во
времени между приходом второго и первого партнеров из пары.

\section{Время пребывания в синхронизаторе} \label{SectTime}

Рассмотрим случайную величину ``время $t$ пребывания в синхронизаторе''.
Пусть в некоторый момент времени $t_0$ в ветви $a$ и $b$ из точки $f$
поступили два партнера одной и той же пары.
Предположим, что в ветви $a$ партнер задержался на время $t_a$ и покинул ее
в момент $t_0 + t_a$.
В ветви $b$ партнер задержался на время $t_b$ и покинул ее
в момент $t_0 + t_b$.
Тогда разница $\tilde{t}$ между временами прихода в синхронизатор партнеров одной и той же
пары, поступивших из ветвей $a$ и $b$,
равна
$\tilde{t} = (t_0 + t_a) - (t_0 + t_b) = t_a - t_b.
$

Обозначим $f_i(t_i)$ плотность распределения вероятности случайной величины
``времени $t_i$ пребывания требования в системе $i$'',
где $i=a \text{ или } i=b$.
Ниже мы покажем, что случайные величины $t_a$ и $t_b$, вообще говоря,
статистически зависимы, а также
проведем количественное исследование этой зависимости. Нам будет удобно сравнивать
реальные характеристики процессов в сети РО с теми, которые получаются в предположении
об отсутствии зависимости между величинами $t_a$ и $t_b$.

Так, для независимых величин $t_a$ и $t_b$,
плотность
распределения $\tilde{f}(\tilde{t})$ случайной величины $\tilde{t}$, определенная
при $\tilde{t} \in (-\infty,+\infty)$, равна свертке:
\begin{equation}
\tilde{f}(\tilde{t}) = \int\limits_{-\infty}^{+\infty} f_b(\tau) f_a(\tau+\tilde{t}) \,d\tau.
\label{EqFTildeTGen}
\end{equation}

Время $t$ ожидания первого партнера в синхронизаторе равно
$$t= |\tilde{t}| = |t_a - t_b|.$$

Плотность $f(t)$ распределения вероятности случайной величины $t$ определена
при $t \in [0,+\infty)$ и равна:
\begin{equation}
f(t)=\tilde{f}(\tilde{t}) + \tilde{f}(-\tilde{t}).
\label{EqFTGen}
\end{equation}

Среднее время $\overline{T}$ пребывания первого партнера в
синхронизаторе выражается интегралом:
\begin{equation}
\overline{T} = \int\limits_{0}^{+\infty} t \,f(t)  \,dt.
\label{EqMeanTGen}
\end{equation}

Отметим, что выражения (\ref{EqFTGen}) и (\ref{EqMeanTGen})
справедливы как при отсутствии, так и при наличии статистической зависимости между $t_a$ и $t_b$.

В Приложении \ref{SectForm} интегралы из выражений (\ref{EqFTildeTGen}) и (\ref{EqMeanTGen})
вычислены при различных значениях $N_a$ и $N_b$.
Оценка для среднего времени $\overline{T}$ пребывания заявки в синхронизаторе,
определяемая формулами
(\ref{EqMeanTGen}, \ref{EqMeanTInf}, \ref{EqMeanT1}),  всегда больше нуля и
зависит от параметров сети РО
($\lambda,\, N_a,\, N_b,\, \mu_a,\, \mu_b$),
где $\lambda$ -- интенсивность входного потока заявок, а
$\mu_i$ -- интенсивность обслуживания в
одном из $N_i$ каналов ветви $i$ ($i=a \text{ или } i=b$).

Как продемонстрировано в Приложении \ref{SectStat}, в численных экспериментах
при всех значениях параметров сети наблюдалось соотношение
$
\overline T > \overline T_{\text{emp}}
$
между эмпирическим значением $\overline T_{\text{emp}}$
среднего времени пребывания в идеальном синхронизаторе,
и предложенным в (\ref{EqMeanTGen})
приближенным выражением $\overline T$ для среднего времени пребывания в
идеальном синхронизаторе. Как видно из таблицы \ref{TableWork},
отклонение не превышало 20\% даже в самом неблагоприятном случае, когда
$N_a=N_b=1$, а $\psi_a$ и $\psi_b$ близки к единице.
Здесь использованы обозначения:
$\psi_i=\lambda/(N_i \mu_i)$.
При других значениях
параметров отклонение на превышает 10\%.

В применении к реальному синхронизатору (в отличие от мгновенно срабатывающего идеального
синхронизатора) выражение
$\overline T_{\text{emp}}$
является оценкой снизу для среднего времени пребывания в синхронизаторе.
Причем эта оценка зависит от параметров сети РО
и не может быть уменьшена за счет повышения производительности реального синхронизатора.

\subparagraph*{Корреляция времен прохождения ветвей $a$ и $b$.}

В реальной сети РО $\{M/M/N_a$; \, $M/M/N_b\}$ потоки заявок,
поступающие на вход синхронизатора из ветвей $a$ и $b$,
коррелируют по причине общего происхождения при разветвлении в
точке $f$ (рис. \ref{PictForkJoinDouble}). Соответственно, (вообще говоря) коррелируют и
времена прохождения ветвей $a$ и $b$ партнерами из одной маркированной пары.
А значит, выражение (\ref{EqFTildeTGen}) для реальной сети РО не всегда является точным.

Очевидно, что при $N_a = N_b = \infty$ корреляция между случайными величинами $t_a$ и $t_b$
отсутствует. Действительно, в этом случае длины очередей в обеих ветвях в точности равны нулю, и
время прохождения каждой ветви равняется
времени обслуживания в отдельном канале обслуживания данной ветви. То есть время прохождения ветви
совершенно не зависит от процессов в соседней ветви сети РО. При этом выражение (\ref{EqFTildeTGen})
становится точным.

По мере уменьшения количества $N_a$ и $N_b$ каналов обслуживания от бесконечности к единице,
в ветвях появляются ненулевые очереди, и времена пребывания в ветвях проявляют все нарастающую
тенденцию к взаимной корреляции. Корреляция достигает максимума при $N_a = N_b = 1$.
При этих значениях параметров отклонение реального распределения $\tilde{f}(\tilde{t})$
от выражения (\ref{EqFTildeTGen}) становится максимальным.

\begin{figure}[htb]
\begin{center}
\includegraphics[width=14cm]{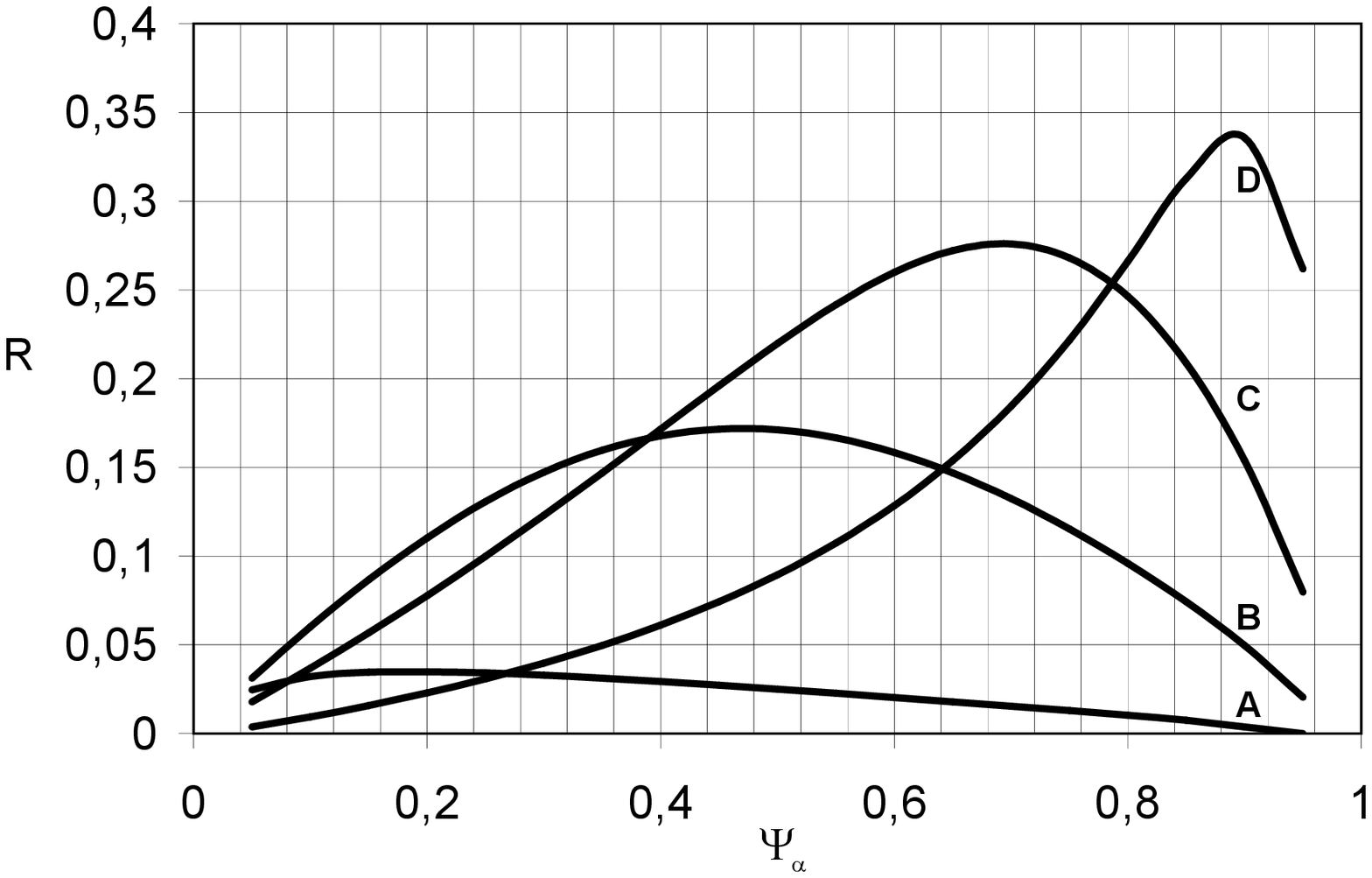}
\end{center}
\caption{Коэффициент корреляции (см. Приложение \ref{SectKolm}) между
временами пребывания в системах $a$ и $b$ для сети $N_a = N_b = 1$
от $\psi_{a}$
при $\psi_{b}=0,05$ (A),
$\psi_{b}=0,35$ (B), $\psi_{b}=0,65$ (C)
и $\psi_{b}=0,90$ (D)
}
\label{PictCorrT}
\end{figure}

Эти качественные рассуждения подтверждаются решением уравнений Колмогорова-Чепмена
(см. Приложение \ref{SectKolm}) в самом
неблагоприятном случае $N_a = N_b = 1$.
На рис. \ref{PictCorrT} показана зависимость коэффициента $R$ корреляции между
случайными величинами
$t_a$ и $t_b$ при $N_a = N_b = 1$ от $\psi_a$ при нескольких значениях $\psi_b$.
Видно, что
коэффициент корреляции заметно отличается от нуля только в тех случаях, когда
интенсивности обслуживания в обеих ветвях примерно одинаковы и при этом лишь ненамного
превышают интенсивность входного потока.

Другим подтверждением приведенных рассуждений являются результаты
численного моделирования (см. Приложение \ref{SectStat})
при произвольных значениях параметров сети РО.
В таблице \ref{TableT} сведены области параметров сети РО, в которых
(согласно критерию Пирсона с уровнем значимости $\alpha=0.01$)
была принята следующая статистическая гипотеза.

\emph{Гипотеза 1.} Наблюдаемое распределение вероятности времени
пребывания в синхронизаторе совпадает с выражениями
(\ref{EqFTGen}, \ref{EqFTInf}, \ref{EqFT1}, \ref{EqFTN})
для $f(t)$ в предположении, что формула (\ref{EqFTildeTGen})
является точной.

\begin{table}[htb]
\caption{
Параметры, при которых Гипотеза 1 принята с уровнем значимости 0.01.
}
\begin{center}
\begin{tabular}{|c|c|c|c|}
\hline
$N_a$  &$N_b$   &$\psi_a$   &$\psi_b$\\\hline
[1, 2]  &[1, 2]  &(0, 0.2]    &(0, 0.2]\\\hline
[3, 5]  &[3, 5]  &(0, 0.5]    &(0, 0.8]\\\hline
[3, 5]  &[3, 5]  &(0, 0.8]    &(0, 0.5]\\\hline
[6, $\infty$)  &[6, $\infty$)  &(0, 0.75]   &(0, 0.75]\\\hline
\end{tabular}
\end{center}
\label{TableT}
\end{table}

Каждая из строк таблицы \ref{TableT} определяет область параметров,
в которых Гипотеза 1 принята с уровнем значимости 0.01.
Видно, что и в произвольном случае
отклонения от приближенного выражения (\ref{EqFTildeTGen})
заметны, только если
интенсивности обслуживания в обеих ветвях примерно одинаковы и при этом лишь ненамного
превышают интенсивность входного потока.

\section{Обсуждение результатов}\label{SectDiscus}

В процитированных во введении работах процесс объединения парных заявок в сетях РО не рассматривался.
В основном авторы вычисляли максимум из времен прохождения всех ветвей сети РО,
не интересуясь процессами, происходящими в точке $S$ и после нее.

Введенное нами представление о новом функциональном элементе, синхронизаторе маркированных
пар, позволило
получить
выражения для функции распределения времени пребывания
заявок в идеальном синхронизаторе.
Характеристики распределений вероятности времени пребывания
заявок в синхронизаторе необходимы при анализе работы всей сети РО.

Эти результаты необходимы и для нахождения распределения
количества требований в синхронизаторе. Так, по закону Литтла,
который справедлив для произвольных систем $G/G/N$ (то есть
и для непуассоновских потоков) \cite{Kleinrock1979},
среднее значение $\rho$ количества заявок в синхронизаторе равно
$$
\rho = \lambda_{\text{in}} \, \overline{T},
$$
где  $\overline{T}$ -- среднее время пребывания в синхронизаторе
(\ref{EqMeanTGen}), а интенсивность потока на входе синхронизатора
$\lambda_{\text{in}} = \lambda$ для стационарного режима в сети РО.

Величина $\rho$, в свою очередь, необходима для
расчета ресурсов, обеспечивающих бесперебойную работу сети РО, а также
для решения задачи об оптимизации этих ресурсов.

Нам приятно поблагодарить
А.В. Колодзея,
А.В. Князева,
К.Ю. Платова,
И.А. Кравченко
и А.Ф. Ронжина
за полезные обсуждения.

\appendix

\section{Гиперэкспоненциальные распределения}\label{SectForm}

Для некоторых фиксированных значений параметров $N_a$ и $N_b$,
рассчитаем
приближенные выражения для функции $f(t)$ распределения
времени $t$ пребывания в синхронизаторе, заданной общей формулой
(\ref{EqFTGen}). Эти выражения будут использованы
при проверке Гипотезы 1 из раздела \ref{SectTime}, а также в
Приложении \ref{SectStat}.

\subparagraph*{Сеть РО $\{M/M/\infty$; \, $M/M/\infty\}$.}

Пусть интенсивность обслуживания в одном канале системы
$a$ равна $\mu_a$, а интенсивность обслуживания в одном канале системы $b$ равна $\mu_b$.
В этом случае плотность $f_i(t_i)$ распределения вероятности случайной величины $t_i$ --
длительности обработки требования в системе $i$ (оно же -- время пребывания в системе $i$)
равна
$$
f_i(t_i)=\mu_i \exp(-\mu_i t_i) \theta(t_i), \quad i=a \text{ или } i=b.
$$
где $\theta(t)$ - функция Хевисайда:
\begin{equation}
\theta(t)=\begin{cases}
1,&\text{если $t\geq0$,}\\
0,&\text{если $t<0$.}\\
\end{cases}
\label{EqFuncHevi}
\end{equation}
Подставляя эти выражения в формулу
(\ref{EqFTildeTGen}), получим плотность $\tilde{f}(\tilde{t})$ распределения
вероятности случайной величины $\tilde{t}$:
$$
\tilde{f}(\tilde{t})=\\
\frac{\mu_a \mu_b}{\mu_a+\mu_b}
\left(
\exp(-\mu_a \tilde{t}) \theta(\tilde{t}) + \exp(\mu_b \tilde{t}) \theta(-\tilde{t})
\right).
$$
Из выражения (\ref{EqFTGen}) следует, что плотность $f(t)$ распределения
вероятности случайной величины $t$ равна
\begin{equation}
f(t)=\frac{\mu_a \mu_b}{\mu_a+\mu_b}
\left(
\exp(-\mu_a t) + \exp(-\mu_b t)
\right).
\label{EqFTInf}
\end{equation}
Таким образом, для сети РО $\{M/M/\infty$; \, $M/M/\infty\}$
время $t$ ожидания в синхронизаторе имеет гиперэкспоненциальное распределение
\cite{Kleinrock1979,Koffman1965}.

Подставляя выражение (\ref{EqFTInf})
в формулу (\ref{EqMeanTGen}), получим среднее время $\overline{T}$ пребывания первого партнера в парном синхронизаторе:
\begin{equation}
\overline{T} =
(\mu_a^2+\mu_b^2)/\left((\mu_a+\mu_b)\mu_a\mu_b\right).
\label{EqMeanTInf}
\end{equation}

\subparagraph*{Сеть РО $\{M/M/1$; \, $M/M/1\}$.}

Для систем $M/M/1$ известно \cite{Kleinrock1979,Koffman1965},
что плотность $f_i(t_i)$ распределения случайной величины $t_i$
(времени пребывания в системе $i$) равна
$$
f_i(t_i)=(\mu_i - \lambda)\exp(-(\mu_i - \lambda) t_i) \theta(t_i),
$$
где $\theta(t_i)$ -- функция Хевисайда (\ref{EqFuncHevi}).
Подставляя эти выражения в формулу (\ref{EqFTildeTGen}),
получим плотность $\tilde{f}(\tilde{t})$ распределения
случайной величины $\tilde{t}$:
$$
\tilde{f}(\tilde{t})=\\
\frac{(\mu_a - \lambda) (\mu_b - \lambda)}{\mu_a + \mu_b - 2 \lambda}
\left(
\exp(-(\mu_a - \lambda) \tilde{t}) \theta(\tilde{t}) + \exp((\mu_b - \lambda) \tilde{t}) \theta(-\tilde{t})
\right).
$$
Из выражения (\ref{EqFTGen}) следует, что плотность $f(t)$ распределения
случайной величины $t$ равна
\begin{equation}
f(t)=\frac{(\mu_a - \lambda) (\mu_b - \lambda)}{\mu_a + \mu_b - 2 \lambda}
\left(
\exp(-(\mu_a - \lambda) t) + \exp(-(\mu_b - \lambda) t)
\right).
\label{EqFT1}
\end{equation}
Таким образом, для сети РО $\{M/M/1$; \, $M/M/1\}$ случайная величина
время $t$ ожидания в синхронизаторе (как и в предыдущем случае)
имеет гиперэкспоненциальное распределение.

Подставляя выражение (\ref{EqFT1})
в формулу (\ref{EqMeanTGen}) найдем среднее время $\overline{T}$ пребывания в синхронизаторе:
\begin{equation}
\overline{T}=
\frac{(\mu_a - \lambda)^2+(\mu_b - \lambda)^2}{(\mu_a+\mu_b - 2 \lambda)(\mu_a - \lambda)(\mu_b - \lambda)}.
\label{EqMeanT1}
\end{equation}

\subparagraph*{Сеть РО $\{M/M/N_a$; \, $M/M/N_b\}$.}

Для общего случая системы массового обслуживания
с произвольным числом параллельных каналов известно
\cite{Koffman1965},
что плотность $f_i(t_i)$ распределения вероятности случайной величины
``время $t_i$ пребывания в системе $i$'' определена при $t_i \in [0,+\infty)$ и равна:
\begin{equation}
f_i(t_i) = \frac{\partial}{\partial\, t_i}
\int\limits_{0}^{t_i}
v_i(\xi)\, W_i(t_i - \xi)\, d\, \xi,
\label{EqFT0}
\end{equation}
где $v_i(x) = \mu_i \exp(-\mu_i x)$
плотность вероятности времени обслуживания в отдельном канале системы $i$;
$W_i(x) = 1 - \tilde{p}_i \exp(-(\mu_i N_i - \lambda)\, x) $
функция распределения времени пребывания в очереди системы $i$;
$\tilde{p}_i$ -- вероятность того, что очередь в системе $i$ непуста:
$$
\tilde{p}_i = p_i^{(0)} \, \frac{(\lambda/\mu_i)^{N_i}}
{N_i!\,(1-\lambda/(\mu_i N_i))};
$$
а $p_i^{(0)}$ - вероятность того, что в системе $i$ нет ни одного требования:
$$
p_i^{(0)} = \left(
\frac{(\lambda/\mu_i)^{N_i}}
{ N_i!\,(1-\lambda/(\mu_i N_i)) }
+ \sum_{k=0}^{N_i-1}\frac{(\lambda/\mu_i)^k}{k!}
\right)^{-1}.
$$
Интегрируя и дифференцируя в выражении (\ref{EqFT0}), получаем:
\begin{equation}
f_i(t_i) =\mu_i\, \left(1 +
\frac{\tilde{p}_i \, \mu_i}{\mu_i\, (N_i-1) - \lambda}
\right) \exp(-\mu_i t_i)
-
\frac{\tilde{p}_i\, \mu_i\, (\mu_i\, N_i - \lambda)}{\mu_i\, (N_i-1) - \lambda}
\exp(-(\mu_i N_i - \lambda)\, t_i).
\label{EqFT01}
\end{equation}

Подставляя выражение (\ref{EqFT01}) в формулы (\ref{EqFTildeTGen}) и (\ref{EqFTGen}),
можно показать, что функция распределения
времени $t$ ожидания в синхронизаторе равна
\begin{equation}
f(t)=
C_1\, \exp(-\mu_a t) +
C_2\, \exp(-\mu_b t) +
C_3\, \exp(-(\mu_a N_a - \lambda) t) +
C_4\, \exp(-(\mu_b N_b - \lambda) t),
\label{EqFTN}
\end{equation}
где коэффициенты $C_1\, \ldots\, C_4$ зависят от параметров
$\lambda,\, N_a,\, N_b,\, \mu_a,\, \mu_b$.
Явный вид этих коэффициентов весьма громоздкий и поэтому здесь не приводится.
Таким образом, и в общем случае сети РО $\{M/M/N_a$; \, $M/M/N_b\}$ случайная величина
``время $t$ ожидания в синхронизаторе''
имеет гиперэкспоненциальное распределение.

\section{Решение уравнений Колмогорова-Чепмена}\label{SectKolm}

В этом разделе рассмотрим
стационарный режим в сети РО с двумя ветвями $\{M/M/1$; \, $M/M/1\}$
(рис. \ref{PictForkJoinDouble}), в каждой из которых имеется по одному
каналу обслуживания $N_a = N_b = 1$, а
на вход сети поступает пуассоновский поток интенсивности $\lambda$.
Состояние ветви $i$ ($i$ принимает значения $a$ или $b$) в некоторый момент времени
будем характеризовать неотрицательным целым числом $q_i$ --
количеством требований, находящихся в ветви $i$.
Состояние обеих ветвей, рассматриваемых совместно,
характеризуется парой чисел $(q_a,\, q_b)$. Вероятность
состояния $(q_a,\, q_b)$ равна $P(q_a,\, q_b)$ и удовлетворяет условию нормировки:
$
\sum_{q_a,\, q_b \ge 0} P(q_a,\, q_b) = 1.
$
Интенсивности
обслуживания в ветвях равны $\mu_{a}$ и $\mu_{b}$.
Напомним, что $\psi_{a}=\lambda/(N_a \mu_a)$ и $\psi_{b}=\lambda/(N_a \mu_b)$.
В стационарном режиме
вероятности $P(q_a,\, q_b)$ не зависят от времени, а
$\psi_{a}<1$ и $\psi_{b}<1$.

Требуется найти совместное распределение вероятности
$P(q_{a},q_{b})$ количества заявок в системах $q_{a}$ и $q_{b}$
в стационарном режиме, для произвольного
фиксированного набора параметров сети РО:
$\lambda,\, \mu_a,\, \mu_b$.

Такое двумерное распределение $P(q_{a},q_{b})$ описывается
системой стационарных уравнений Колмогорова-Чепмена:

\noindent
при $q_{a}=q_{b}=0$:\qquad $\lambda P(0,0)=\mu_{a}P(1,0)+\mu_{b}P(0,1)$,\\
при $q_{b}>0$:\qquad
$(\lambda+\mu_{b})P(0,q_{b})=\mu_{a}P(1,q_{b})+\mu_{b}P(0,q_{b}+1)$,\\
при $q_{a}>0$:\qquad
$(\lambda+\mu_{a})P(q_{a},0)=\mu_{a}P(q_{a}+1,0)+\mu_{b}P(q_{a},1)$,\\
при $q_{a} q_{b}>0$:\qquad
$(\lambda+\mu_{a}+\mu_{b})P(q_{a},q_{b})=\lambda P(q_{a}-1,\,q_{b}-1)+\mu_{a}P(q_{a}+1,\,q_{b})+\mu_{b}P(q_{a},\,q_{b}+1)$.

К сожалению, точное решение этих уравнений в данном случае неизвестно.
Поэтому поставленная задача решалась численно итерационным методом.
В качестве начального приближения $P_{0}(q_{a},q_{b})$ использовалось распределение,
соответствующее случаю, когда на вход систем
$a$ и $b$ поступают два независимых пуассоновских потока интенсивности $\lambda$:
$$
P_{0}(q_{a},q_{b})=(1-\psi_{a})\psi_a^{q_a}(1-\psi_b)\psi_b^{q_{b}}.
$$
Каждая следующая итерация вычислялась по формулам:
$$
P_{n+1}(0,0)=(1-\gamma)P_{n}(0,0) +
\gamma\frac{\mu_{a}P_{n}(1,0)+\mu_{b}P_{n}(0,1)}{\lambda},
$$
при $q_{b}>0$\ :
$$
P_{n+1}(0,q_{b})=
(1-\gamma)P_{n}(0,q_{b})+
\gamma\frac{\mu_{a}P_{n}(1,q_{b})+\mu_{b}P_{n}(0,q_{b}+1)}{\lambda+\mu_{b}},
$$
при $q_{a}>0$\ :
$$
P_{n+1}(q_{a},0)
=(1-\gamma)P_{n}(q_{a},0)+
\gamma\frac{\mu_{a}P_{n}(q_{a}+1,0)+\mu_{b}P_{n}(q_{a},1)}{\lambda+\mu_{a}},
$$
при $q_{a}q_{b}>0$\ :
$$
P_{n+1}(q_{a},q_{b})
=(1-\gamma)P_{n}(q_{a},q_{b})+
\gamma\frac{\lambda P_{n}(q_{a}-1,q_{b}-1)+
\mu_{a}P_{n}(q_{a}+1,q_{b})+\mu_{b}P_{n}(q_{a},q_{b}+1)}
{\lambda+\mu_{a}+\mu_{b}}.
$$

Параметр $0<\gamma \le 1$ регулировался в процессе итераций
(фактически, использовалось два значения: $\gamma=1$ и $\gamma=0.1$).
При вычислениях использовалась матрица значений $P_{n}(q_{a},q_{b})$ размером
$N\times N$, где $N=190$. Предполагалось, что $P_{n}(q_{a},q_{b})=0$,
как только хотя бы один из аргументов $q_i$ выходит за границы отрезка
[0,\, $N-1$].

В качестве критерия близости $P_{n}$ к истинному значению использовалась
малость трех величин:
$$
D_{1}=\sum\limits_{h=0}^{N-1} \left|
p_a(h) - {\sum\limits_{l=0}^{N-1} P_n(h,l)}
\right|,
$$
$$
D_{2}=\sum\limits_{h=0}^{N-1} \left|
p_b(h) - {\sum\limits_{l=0}^{N-1} P_n(l,h)}
\right|,
$$
$$
D_{3}=\sum\limits_{h=0}^{N-1}
\sum\limits_{l=0}^{N-1} \left| P_{n+1}(h,l) - P_{n}(h,l)\right|.
$$
где  $p_i(m)=(1-\psi_i)\psi_i^{m}$ --
безусловная вероятность того,
что количество заявок $q_i$ в ветви $i$ равно $m$.

После каждой итерации все значения $P_{n+1}$ нормировались,
то есть делились на сумму
$$\sum\limits_{h=0}^{N-1} \sum\limits_{l=0}^{N-1} P_{n+1}(h,l).$$

Использовался следующий алгоритм выбора параметра $\gamma$:
сначала выполнялись итерации при $\gamma=1$, затем, когда параметр $D_{3}$
переставал уменьшаться, выполнялись $10^3$ итераций при $\gamma=0.1$. Затем
снова выполнялись итерации при $\gamma=1$.

Примененный итерационный метод обеспечивает невозрастание
абсолютной величины максимальной
(по всем $0 \le h \le N-1$ и $0 \le l \le N-1$)
из абсолютных погрешностей $\Delta_{n}(h,l) = P_{n+1}(h,l) - P_{n}(h,l)$
приближений для вероятностей $P(h,l)$, поскольку
$$
|\Delta_{n+1}(h,l)|=
$$
$$
=\left|(1-\gamma)\Delta_{n}(h,l)+
\gamma\frac{\lambda \Delta_{n}(h-1,l-1)+
\mu_{a}\Delta_{n}(h+1,l)+\mu_{b}\Delta_{n}(h,l+1)}
{\lambda+\mu_{a}+\mu_{b}}\right|\le
$$
$$
=\max_{h,l}(|\Delta_{n}(h,l)|).
$$

Наблюдаемое в процессе итераций быстрое уменьшение
невязки системы уравнений при всех исследованных комбинациях
параметров сети РО позволяет предположить, что применение нормировки
после итерации приводит к изменению знаков абсолютных погрешностей,
что в свою очередь превращает нестрогое неравенство, характеризующее
уменьшение максимума абсолютной ошибки, в строгое, и обеспечивает сходимость метода.
Отметим, однако, что строгим доказательством сходимости
итераций к решению системы уравнений Колмогорова-Чепмена
для произвольной комбинации параметров
авторы не располагают.

В рамках приведенного численного метода были получены
приближенные решения для $P(q_{a},q_{b})$ при
$0.05\le q_{a},q_{b} \le 0.9$. Характерные значения параметров $D_{1},D_{2},D_{3}$,
при прекращении вычислений приведены в таблице \ref{TableD}.

\begin{table}[htb]
\caption{Характерные значения величин при остановке итераций}
\begin{center}
\begin{tabular}{|c|c|c|} \hline
$\psi_{a}=\psi_{b}$
&$D_{1}=D_{2}$
&$D_{3}$ \\  \hline
$0.05$ & < $10^{-11}$    & < $10^{-11}$   \\  \hline
$0.5$  & < $10^{-11}$    & < $10^{-11}$   \\  \hline
$0.7$  & $0.12\cdot10^{-8}$  & $0.2\cdot10^{-10}$ \\  \hline
$0.9$  & $0.76\cdot10^{-3}$  & $0.53\cdot10^{-5}$ \\  \hline
\end{tabular}
\end{center}
\label{TableD}
\end{table}

Полученные приближенные решения для  $P(q_{a},q_{b})$ позволяют
количественно охарактеризовать наличие корреляции между процессами
в ветвях сети РО $\{M/M/1$; \, $M/M/1\}$.
На рисунке \ref{PictCorrT} показана зависимость коэффициента корреляции
между временами нахождения партнеров одной пары в ветвях сети РО от
$\psi_{a}$ для четырех значений $\psi_{b}$:
0.05;\,
0.35;\,
0.65 и 0.90. Как видно из графиков,
заметный рост коэффициента корреляции, так же как и
заметное изменение условной вероятности, наблюдается в случаях, когда
интенсивности обслуживания в обеих ветвях сети РО
примерно одинаковы ($\psi_{a}=\psi_{b}$).

\section{Численные эксперименты}\label{SectStat}

Для моделирования случайных процессов в сети РО (рис. \ref{PictForkJoinDouble})
применялся входной поток из $10^5$ заявок с экспоненциальным распределением временных
интервалов, полученный с использованием датчика случайных чисел.

\begin{table}[htb]
\caption{Проверка Гипотезы 1 (с уровнем значимости $\alpha = 0.01$)
и наблюдаемые значения $\Delta \overline T/\overline T$.}
\begin{center}
\begin{tabular}{|c|c|c|c|c|c|c|c|}
\hline
\multicolumn{8}{|c|}{$\lambda=0.3,\,N_{a}=1,\,N_{b}=1$}\\\hline
$\psi_{a}$&0.75 &0.75   &0.75   &0.75   &0.1    &0.1    &0.375\\\hline
$\psi_{b}$&0.75 &0.5    &0.25   &0.2    &0.2    &0.1    &0.375\\\hline
$\chi^{2}$&\textbf{1551}    &\textbf{440}   &\textbf{112}
&\textbf{115}   &45     &48     &\textbf{760}\\\hline
$\Delta \overline T/\overline T \times 100\%$ &17.5 &7.3 &0.6 &0.3 &1.3 &1.4 &8.7\\\hline
\multicolumn{8}{|c|}{$\lambda=1.5,\,N_{a}=3,\,N_{b}=5$}\\\hline
$\psi_{a}$&0.83 &0.91   &0.83   &0.83   &0.625  &0.5    &0.25\\\hline
$\psi_{b}$&0.3  &0.6    &0.75   &0.83   &0.6    &0.5    &0.25\\\hline
$\chi^{2}$&33   &\textbf{142}   &\textbf{674}   &\textbf{895} &\textbf{84} &29 &35.5\\\hline
$\Delta \overline T/\overline T \times 100\%$ &1.4 &7.3 &9.4 &9.7 &2.0 &0.9 &0.04\\\hline
\multicolumn{8}{|c|}{$\lambda=2,\,N_{a}=8,\,N_{b}=8$}\\\hline
$\psi_{a}$&0.5  &0.83   &0.93   &0.83   &0.9    &0.42   &0.625\\\hline
$\psi_{b}$&0.36 &0.625  &0.71   &0.83   &0.93   &0.83   &0.5\\\hline
$\chi^{2}$&33.2 &44     &\textbf{142}   &\textbf{227}   &\textbf{480}   &34.8   &42\\\hline
$\Delta \overline T/\overline T \times 100\%$ &0.2 &1.8 &6.9 &5.5 &7.9 &0.5 &0.2\\\hline
\end{tabular}
\end{center}
\label{TableWork}
\end{table}

Проверялась статистическая Гипотеза 1 из раздела \ref{SectTime}
о совпадении эмпирического распределения времени пребывания в синхронизаторе
с гипотетическим по критерию $\chi^2$ Пирсона.
В качестве гипотетических распределений использовались приближенные выражения
(\ref{EqFTGen}, \ref{EqFTInf}, \ref{EqFT1}, \ref{EqFTN}).
Область определения плотности распределения разбивалась на 30 отрезков равной вероятности.
При многих комбинациях значений параметров сети РО
($\lambda,\, N_a,\, N_b,\, \psi_a,\, \psi_b$) рассчитывались значения
критерия $\chi^2$ Пирсона.
Принимался уровень значимости $\alpha = 0.01$.
При этом граничное значение критерия Пирсона
$\chi_0^2 = 49.6$.

В таблице \ref{TableWork} приведены
величины $\chi^{2}$ для некоторых значений параметров сети.
Жирным шрифтом выделены величины, при которых
Гипотеза 1 была отвергнута:
$\chi^2 > \chi_0^2$.
Аналогичные данные были использованы для построения таблицы \ref{TableT}.
Кроме того, в таблице \ref{TableWork}
приведена относительная погрешность
$\Delta \overline T/\overline T=(\overline T - \overline T_{\text{emp}})/\overline T$,
показывающая, как отклоняется
эмпирическое среднее время $\overline T_{\text{emp}}$ от
приближения $\overline T$, определяемого формулами
(\ref{EqFTildeTGen}), (\ref{EqMeanTGen}), (\ref{EqMeanTInf}) и (\ref{EqMeanT1}).

\end{document}